# Charge Density Wave Order in Superconducting Topological Insulator Nb$_x$-Bi$_2$Se$_3$

**Yanan Li[1], Christian Parsons[1], Sanath Ramakrishna[2], Anand Dwivedi[1], Nelson William[2], Ryan Baumbach[2], Marvin Schofield[1], Arneil Reyes[2], and Prasenjit Guptasarma[1*]**

[1]Department of Physics, University of Wisconsin-Milwaukee, Milwaukee, Wisconsin 53211, USA
[2]National High Magnetic Field Laboratory/Florida State University, Tallahassee, Florida 32310, USA

**Abstract**

Spontaneously broken symmetry in the ground state of Bi-based topological materials can lead to promising Topological superconductors, such as Cu-Bi$_2$Se$_3$, Sr-Bi$_2$Se$_3$ and Nb-Bi$_2$Se$_3$. In recent studies, coexistence of multiple Fermi surface features and superconductivity was found in Nb-Bi$_2$Se$_3$. However, the resultant mutliple Fermi surface--charge density wave has not been experimentally reported yet. In this paper, we report possible evidence of co-occurence of a charge density wave (CDW) ground state and a superconducting (SC) ground state in Nb intercalated Bi$_2$Se$_3$. An intercalation induced Periodic Lattice Distortion appears to help stabilize the CDW state, possibly assisting in the formation of 1D chains and rendering electronic and phonon anisotropy to this intriguing system.

Keywords: Superconductivity, Charge Density Wave, Fermi Surface, Periodic Lattice Distortion, Nb$_x$Bi$_2$Se$_3$, Topological Insulator, Intercalation

## 1. Introduction

The interplay between charge density wave (CDW) order and unconventional superconductivity has been of intense interest in Condensed Matter Physics,[1-4] because both orders use electron-phonon/electron interactions and feature broken symmetries at the ground state. Well-known systems which display both a CDW state and an unconventional superconducting state include several layered dicalcogenides,[5-13] where the topology of Fermi surface plays an important role for the occurrence of these two states. Recent reports of broken symmetry in topological insulators such as Bi$_2$Se$_3$, intercalated with Cu, Nb, and Sr, have reinvigorated an interest in examining such intercalated chalcogenides in the context of topological superconductivity.[14-16] Proposals from first principles electronic structure calculations, recently supported by neutron diffraction measurements, suggest that unconventional electron pairing in superconducting Cu- and Sr- intercalated Bi$_2$Se$_3$ might be driven by a singularity in electron-phonon interactions arising from strong Fermi surface Nesting at long wavelength, along $\Gamma_z$ direction.[17,18] With the observation of Strong Fermi surface Nesting and multiple Fermi pockets,[19] it is no surprise that charge density

wave order can be also expected in the intercalated Bi$_2$Se$_3$ system, such as in Nb$_x$Bi$_2$Se$_3$.[20]

In the odd-parity p-wave superconductor Nb$_x$Bi$_2$Se$_3$,[21] symmetry-breaking features such as nematic order appears to coexist with superconducting order, and seem to show up both below and above superconducting T$_c$.[22] Nb$_x$Bi$_2$Se$_3$ displays a complex electronic band structure, with a Fermi pocket off the Brillouin Zone center in addition to the main ellipsoidal Fermi surface characteristic of this class of systems.[20] Fermi surface nesting resulting from such multiple surfaces is expected to yield concomitant density waves in Nb$_x$Bi$_2$Se$_3$.[15,20] Previous work on compounds such as Bi$_2$Te$_3$ and Bi$_2$Se$_3$ had revealed a hexagonal Fermi surface with two flat segments facing each other, and separated by 2k$_F$ along the Γ-K direction leading to predictions of strong nesting and density wave order in such systems as well.[23,24] Experimental observations providing hints of a density waver order, or a resultant gap, include Angle Resolved Photoemission Spectroscopy in Bi$_2$Te$_3$ and Bi$_2$Se$_3$, Scanning Tunnelling Microscopy imaging and first-principles calculations at antiphase grain boundaries of Bi$_2$Se$_3$ thin films, and electron diffraction from single crystals of Cu$_x$- $Bi_2Se_3$.[25-28] However, direct experimental observations of a density wave gap, or a transition to density wave order in superconducting topological insulators using methods such as resistivity, have been lacking. This has hampered, among other things, discussions of relationships between charge order, symmetry breaking, and superconductivity in these systems.

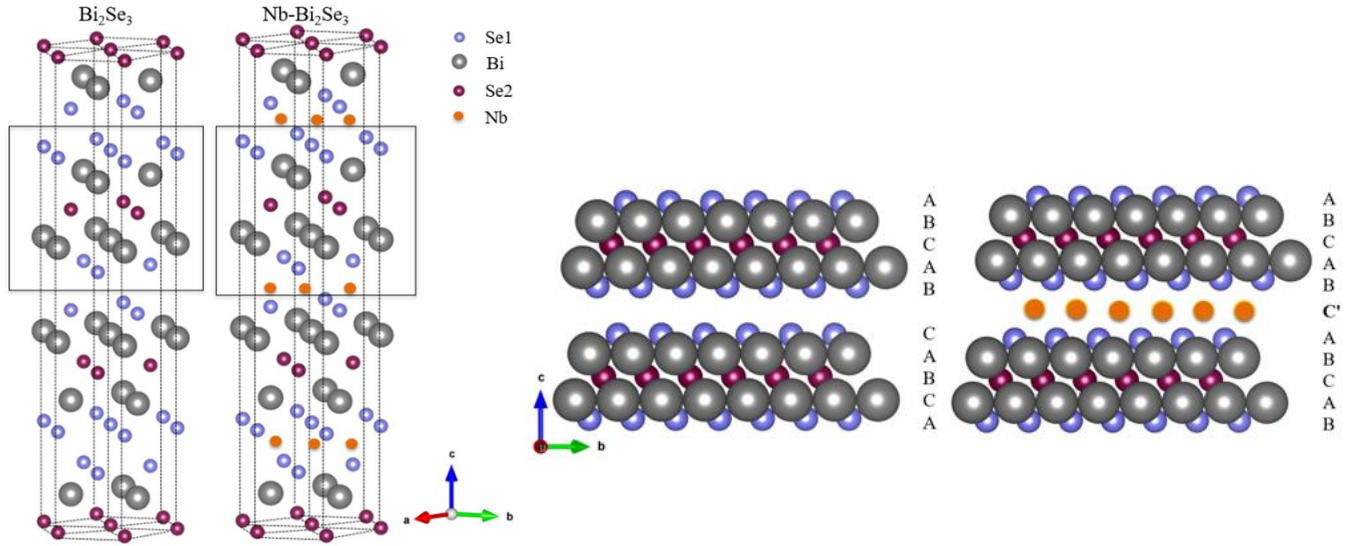

**Fig.1.** Crystal structure of $Nb_xBi_2Se_3$ and ABC stacking symmetry: (a) Crystal structures of $Bi_2Se_3$(left) and Nb intercalated $Bi_2Se_3$ (right). Inside the Quintuple Layer structure of $Bi_2Se_3$ (Square shape), blue solid ball represents Se1 (outer layer Se site), grey solid ball represents Bi, red solid ball Se2 (inter layer Se site) and in the case of $Nb_xBi_2Se_3$, orange ball represents intercalated Nb. (b) ABC stacking along c-axis for $Bi_2Se_3$(left), with stacking order along c-axis as –A(Se1)–B(Bi)–C(Se2)–**A(Bi)–B(Se1)–C(Se1)–A(Bi)** – B(Se2)–C(Bi)–; while ABC stacking falls with Nb intercalated into Van der Waals gap of $Bi_2Se_3$(right), where C'(Nb) breaks the original C(Se1) site, with c-axis now stacking as –A(Se1)–B(Bi)–C(Se2)–**A(Bi)–B(Se1)–A(Se1)–B(Bi)** –C(Se2)–A(Bi)–.

In this article, we reveal that a Charge Density Wave (CDW) order forms at high temperature in $Nb_xBi_2Se_3$ and possibly co-exists with superconductivity (SC) at low temperature. Our observations suggest that an Incommensurate CDW at room temperature orders into a commensurate CDW near 150-200K, far above the superconducting $T_c$ around 3K. In principle, a density wave with 2-fold symmetry could independently give rise to nematic order. Several chalcogenides, with hexagonal crystal symmetry, display spontaneous 2-fold nematic order.[21,29,30] In this work we are unable to rule out whether such order pre-existing CDW in the background. Previous studies by Cho,[22] Hecker[31] suggest that nematic order of similar symmetry might exist both above and below superconducting $T_c$ in superconducting chalcogenides such as $Nb_xBi_2Se_3$ and $Cu_xBi_2Se_3$. It is also interesting to question whether both ground states (superconducting and CDW) draw upon electron-phonon interactions of the same symmetry. We grew different Nb concentration doped $Bi_2Se_3$ by a self-flux growth method followed by high temperature (650C) quenching. We identified the CDW state of the crystals from our structural and electronic measurements. By performing Selected Area Electron Diffraction (SAED), we found evidence of Incommensurate CDW, shown as superlattice and "diffused streaks".[27,32-35] Electronically, we performed Four-Probe Resistivity measurement with temperature dependence, where the metal-to-insulator-like transitions, the indication of an opening of energy gap, were seen in most of our samples. The coexistence of CDW and SC were seen in the transport measurements. Additionally, superconducting $T_c$ was also measured by magnetization as a function of temperature. Furthermore, CDW phase transition was further examined by $^{209}$Bi Nuclear Magnetic Resonance frequency sweep measurement, where each individual peak of spin-9/2 $^{209}$Bi nuclear in $Nb_{0.05}Bi_2Se_3$ shows a homogenous broadening of Full Width at Half Maximum (FWHM) with lowering of temperature.

## 2. Results

Fig.1 (a) displays the crystal structures of host $Bi_2Se_3$ (left) and Nb intercalated $Bi_2Se_3$ (right). Bulk host crystal $Bi_2Se_3$ belongs to $D_{3d}^5$ R$\underline{3}$m space group.[36] The supercell of $Bi_2Se_3$ can be considered as a hexagonal layered structure, with Quintuple layers (QL-Se1-Bi-Se2-Bi-Se1-) of atoms stacked along the trigonal axis. Inside of each Quintuple layer, Se has two possible positions, with Se1 representing the outer layer Se atoms and Se2 representing the inner layer Se atoms.

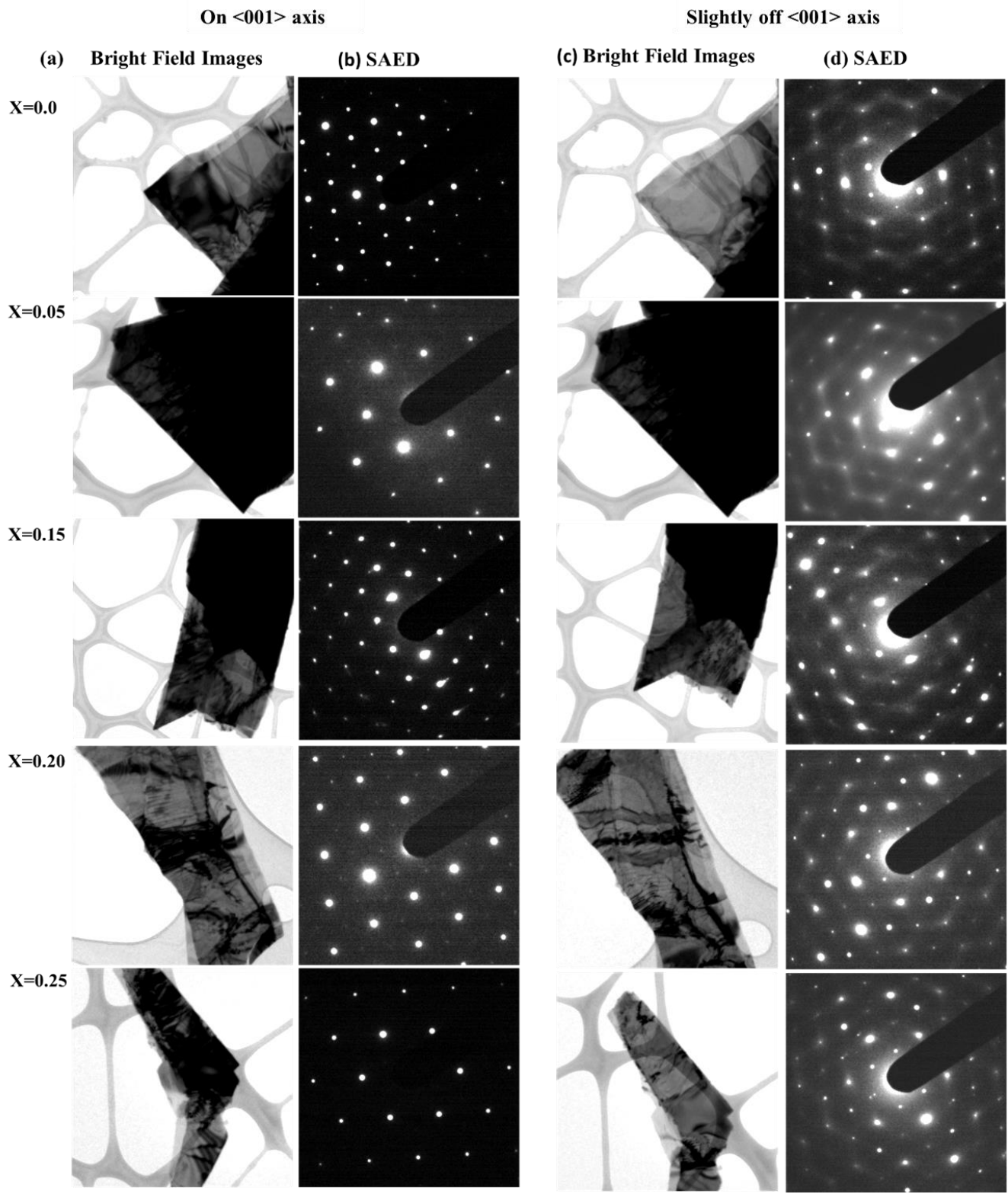

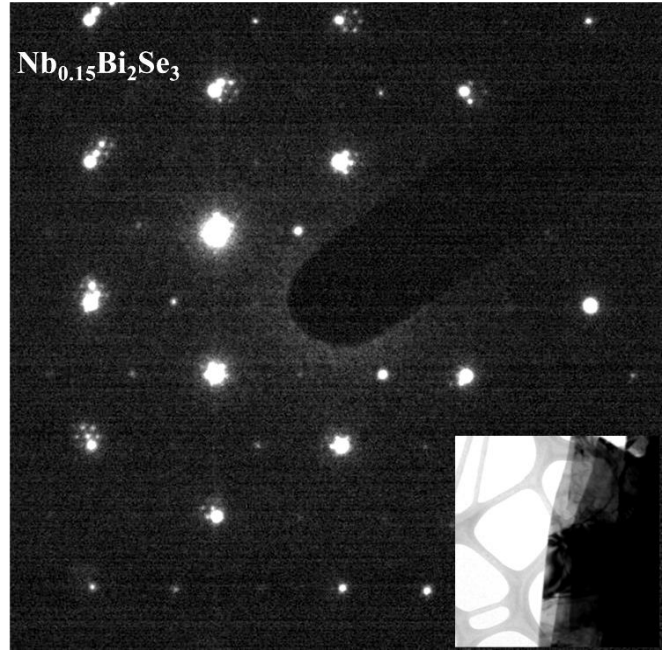

**Fig. 2**. Bright field Transmission Electron Microscopic Images and Selected Area Electron Diffractogram (SAED) for $Nb_xBi_2Se_3$ with x=0.00, 0.05, 0.15, 0.20 and 0.25. Each row in the figure corresponds to a specific value of x, as shown. Each column in the figure corresponds to a different type of image, as described here. (a) Bright field images on <001> zone axis; (b) SAED on the same <001> zone axis; (c) Bright field image with a slight off-zone-axis tilt of less than 5 degree; (d) SAED with the same slight off-zone-axis tilt of less than 5 degree, as in (c). Figure (e) shows on <001> zone axis SAED results from a second piece of the x=0.15 sample, with the corresponding bright field image as an inset . Note six small, weak satellite spots near each $Bi_2Se_3$ Bragg spot, indicating the formation of a superlattice.

Bi only has one site. The coupling between each adjacent atom planes within a QL is a strong covalent bond, while between the QLs are weak van der Waals interactions, which allows the crystal to be easily cleaved along inner QL plane. Each atomic layer has three possible positions along the c-axis of the crystal, stacked in A(Se1)–B(Bi)–C(Se2)–A(Bi)–B(Se1)–C(Se1)–A(Bi) –B(Se2)–C(Bi)– ABC order,[36] as shown in Fig. 1(b)-$Bi_2Se_3$(left). With foreign atoms' intercalation, the intercalants (in this case Nb) tend to go in-between the weak van der Waals gap, which breaks the ABC stacking symmetry of the $Bi_2Se_3$ host structure. Now with C' (Nb) replacing the C(Se1) stacking, the new stacking order becomes –A(Se1)–B(Bi)–C(Se2)–A(Bi)–B(Se1)–A(Se1)–B(Bi)–C(Se2)–A(Bi)–, shown in Fig.1(b) Nb-$Bi_2Se_3$(right). This ABC stacking faults further leads to a lattice disorder observed as Periodical Lattice Distortion in our following Selected Area Electron Diffraction (SAED) measurements, shown as Fig.2.

Fig.2 displays Electron Diffraction images for bright field images on single crystal pieces with x=0.00, 0.05, 0.15, 0.20 and 0.25, and Select Area Electron Diffractions along <001>-zone axis and tilted slightly off the zone axis. The strong diffraction contrast with layered structure can be seen in Fig. 2 (a), indicating good crystal quality and single crystal features. In Fig. 2 (b), besides the $Bi_2Se_3$ host structure, the normally forbidden 2/3 of the reciprocal lattice diffraction spots are also observed with weak Diffraction features. Usually these weak diffraction spots are generated by ABC stacking fault where along the c-axis, Nb or Bi intercalation changes the local structual stability and breaks the original structure symmetry, as illsurated in Fig.1(a) and (b).[27,32,37,38] To further examine the reason behind these extra diffraction spots and to reduce dynamic scattering in considering of the strong diffraction on <001> zone axis orientation,[32] we tilted each of the crystals by less than 5 degrees, shown in Fig 2 (c) and (d). In this case we not only still see these extra diffraction spots, we also observe diffuse "streaks", which are located between the reflections of the Bragg spots. From earlier studies,[27,32,33-35] we know that such diffuse streaks are an indication of Incommensurate Charge Density Wave related to structural lattice disorder, which leads us to look for possible ordered Charge Density Wave at lower temperature. Note that on a different piece of x=0.15 shown as Fig 2 (e), we additionally observe satellite spots around each of the strong diffraction spots. A similar discussion of I-CDW/CDW related with satellite spots can also be found in a study of zero metal intercalated $Bi_2Se_3$ by Koski et al,[27,38] in which different intercalating atoms result in different I-CDW/CDW superlattices.

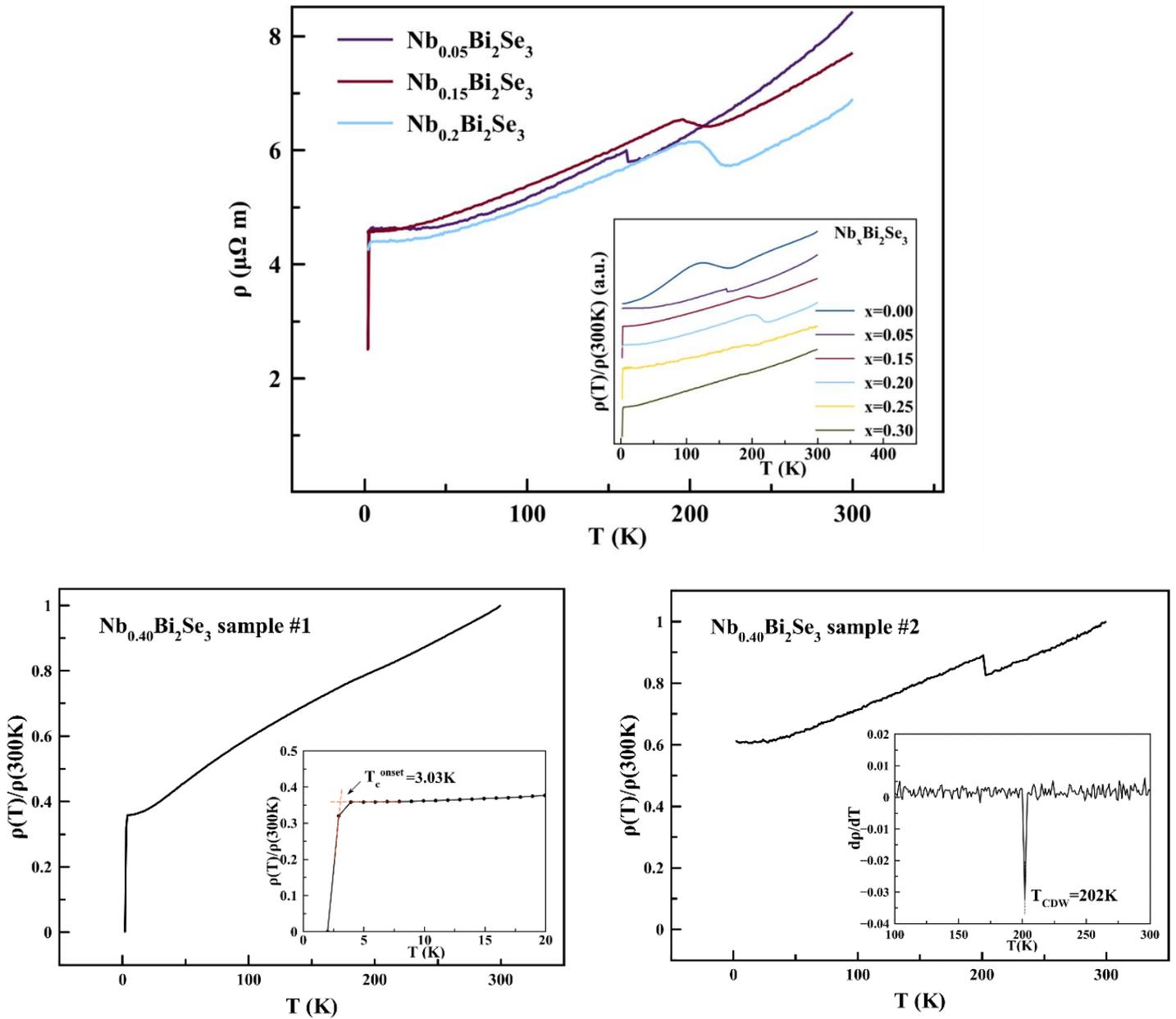

Figure 3. (a) Temperature dependence of resistivity ρ(T) for Nb$_x$Bi$_2$Se$_3$ (x=0.05, 0.15, 0.20) crystal. Inset: ρ (T)/ρ (300K) as a function of T for Nb$_x$Bi$_2$Se$_3$ (x=0.00, 0.05, 0.15, 0.20, 0.25 and 0.30). Note Charge Density Wave transition temperatures T$_{CDW}$ in the temperature range 140K - 200K. Superconducting transitions are observed between 0.0K - 3.9K. (b),(c) are two different pieces of x=0.40. Interestingly, sample #1 in (b) displays only a SC transtion, whereas sample #2 in (c) displays only a CDW transition. Inset graph of (b) is T$_c$$^{onset}$, taken from the intersection of lines from the normal resistence curve and the linear drop of the superconduting state. Inset graph of (c) is T$_{CDW}$, taken from the lowest inflection point of dρ(T)/dT.





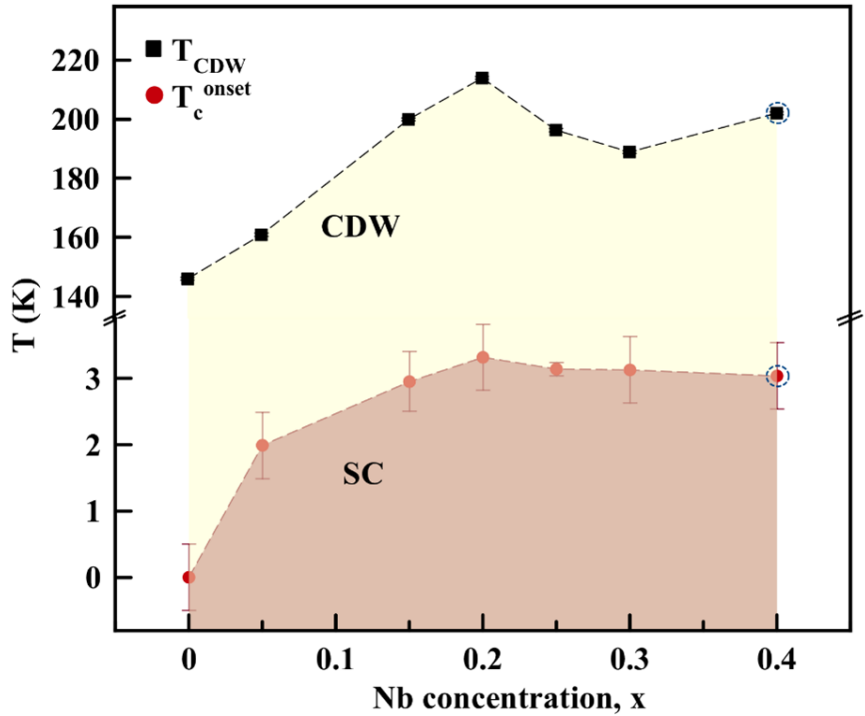

Figure 4. Phase diagram of $Nb_xBi_2Se_3$ showing temperature of transition varying with nominal Nb concentration x and superconducting onset. Charge Density Wave phase (black square) and Superconducting Phase (red dot) serve to differentiate the two phases. Reported density wave transition temperatures $T_{CDW}$ correspond to the lowest inflection point in $d\rho(T)/dT$ curves for CDW, same as 3(a) inner. Superconducting transition $T_c^{onset}$ corresponds to the superconducting onset for each sample, same as 3(b) inner. Bule circles for the x=0.4 Nb concentration sample are used to indicate that the SC and CDW were not found to co-exist but were instead taken from different pieces. In all of the remaining samples, SC and CDW were measured from the same sample.

Fig3(a) displays zero-field resistivity on the ab-plane of $Nb_xBi_2Se_3$ (x=0.05, 0.15 and 0.20), measured on freshly cleaved surfaces, as a function of temperature varying from 2K to 300K. The inset of Fig 3(a) displays $\rho(T)/\rho(300K)$ for several different concentrations x of Nb-intercalated $Bi_2Se_3$ crystals. Note the metal-to-insulator-like transitions observed on each of the crystals, signalling the opening of an energy gap reminiscent of a density wave transition. Based on our electron diffraction results displayed in Figure 2, indicative of an I-CDW, we conclude here that the features observed in resistivity in Figure 3(a) correspond to a transition from a higher temperature I-CDW to a CDW. In other words, the I-CDW "locks" into an ordered charge density wave (CDW) state.[32,34,37,39] As is clear from figures 3(a), the CDW transitions occur in the range of 140K - 200K for all Nb-intercalated samples reported here. Transition temperature in Figure 4(a) were determined from $\rho(T)$ measurements by examining the lowest inflection point in $d\rho/dT$ for CDW and the Superconducting onset as $T_c$. For x>0.00, note a sharp drop of resistivity at lower temperature, 1.9K - 3.4K, which are similar to those previously reported by Kobayashi et al.[40] Additionally, we report magnetic susceptibility versus temperature in Figure 5.

Fig. 4 shows observed CDW and SC transition temperatures on our single crystals of $Nb_xBi_2Se_3$ with varying x. Except for the samples with x = 0.0 and 0.4, all samples displayed here showed both a CDW transition and a SC transition on the same piece. No direct competition between CDW and superconducting states was observed from the phase diagram at x≤0.20. In fact, zero-field transport measurements show that up to x=0.2, both $T_c^{onset}$ and $T_{CDW}$ are increasing with the increasing Nb concentration. The coexistence of the two phases makes the samples more interesting. We attribute the coexistence of the two phases to our growth conditions that caused intercalation and structural disorder. The onset of SC and CDW for x=0.40 sample are from two different pieces, sample#1 and #2. This separation of CDW and SC can be an





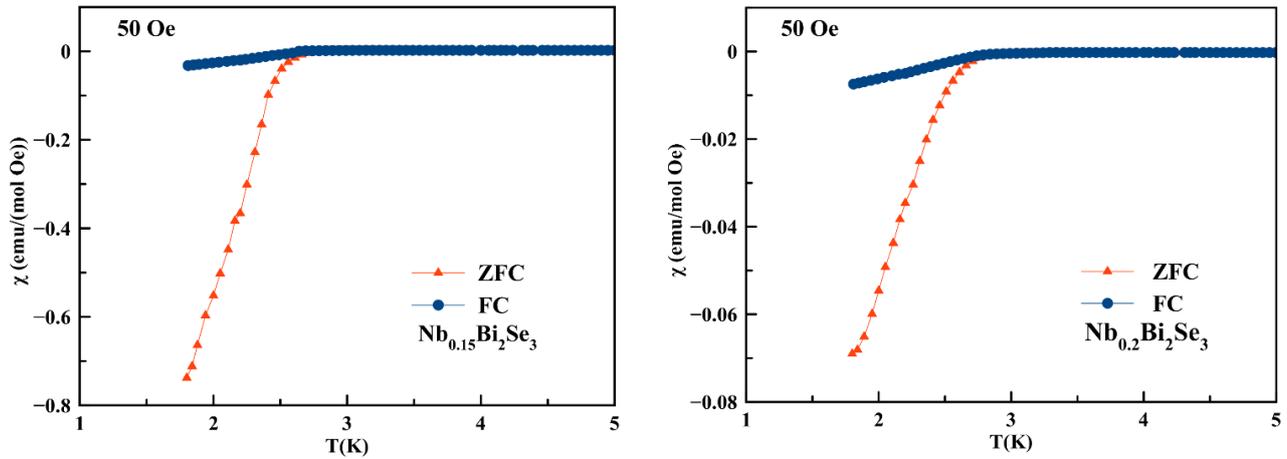

Figure 5 Temperature dependent DC magnetic susceptibility of single crystals of $Nb_{0.15}Bi_2Se_3$ and $Nb_{0.20}Bi_2Se_3$. (a) and (b) Under a nominal applied magnetic field of 50 Oe, ZFC and FC susceptibility for $Nb_{0.15}Bi_2Se_3$ and $Nb_{0.2}Bi_2Se_3$ from 1K to 5K.

indication of the competition between the two-phase transitions, or the impurity suppressed the CDW state.

Fig.5 displays DC magnetic susceptibility in zero-field cooled (ZFC) and field-cooled (FC) modes, with applied field of 50Oe during field cooling and during measurement in the warm-up cycle. The single crystals were mounted inside plastic straws, with c-axis parallel to the applied magnetic field and with plugged-in quartz wool to stabilize the sample inside the straw. Due to limitations of size and shape of the straw and the crystal, the x=0.2 sample had to be mounted slightly off c-axis center. Note superconducting transitions for (a) $Nb_{0.15}Bi_2Se_3$, and (b) $Nb_{0.2}Bi_2Se_3$ with $T_c$ onsets of 2.8K and 3.5K, respectively. These $T_c$ values agree with the onset temperature measured by resistivity, of 2.95K and 3.32K, respectively.

Fig 6. displays the [209]Bi NMR measurement on crystal $Nb_{0.05}Bi_2Se_3$, with the applied static magnetic field magnitude H=10.57T (our sample at non-superconducting state, superconducting $H_{c2}$<3T)[60] and orientation parallel to the c-axis of the crystal. As shown in Fig.6 (a), frequency sweep spectra (from 72 to 74MHz) were taken at temperature 4K, 150K and 200K, which are below, near and above $T_{CDW}$ (=161K for $Nb_{0.05}Bi_2Se_3$). With fitting of Gaussian functions, total of 9 peaks from spin-9/2 of [209]Bi were shown in each temperature's spectra. The full-width at half-maximum (FWHM) of individual peak line shape vs peak number were plotted to quantify the lines broadening as a function of temperature, as shown in Fig. 6(b). Except for the third satellite transition on the left side, all the other peaks show a feature of homogeneous broadening (~30%) with lowering of

the temperature to below $T_{CDW}$. Besides the individual peak broadening, the satellite transitions are also increasingly broadened and become more asymmetric as their location increases from the central transition. Same observation has been discussed in Willson's dissertation-Figure 5.8 for spin 9/2 [93]Nb in $NbSe_2$,[41] where stated this is the evidence of the CDW phase because the quadrupolar splitting is an electric field gradient induced effect and would most certainly be affected by a long-range charge order.

## 2. Discussion

### 2.1.1 NMR

NMR is a magnetic local probe to detect the microscopic information about the spin dynamics, chemical and structure natures of materials through coupling of a specific nuclei with its local environment. [42,43] In NMR, both the position and shape of the absorption line contain structural and dynamical information concerning the nucleus with its local environment.[43]

Contributions to NMR line shape have homogenous and inhomogeneous factors, where homogenous line shape usually found in solids and single crystals can be fitted with Gaussian, while the inhomogeneous line shape has separate parts that originated from separate contributions. The line broadening can be either due to inhomogeneous line shape or due to a wide range of homogenous factors such as phase transitions, where CDW is one of the specific examples. In the case of CDW, below the transition, the periodic modulation of the charge density can lead to periodic modulation of the electric-field-gradient (EFG) tensor. The symmetry effects are particularly spectacular on the electric field gradient since the NMR





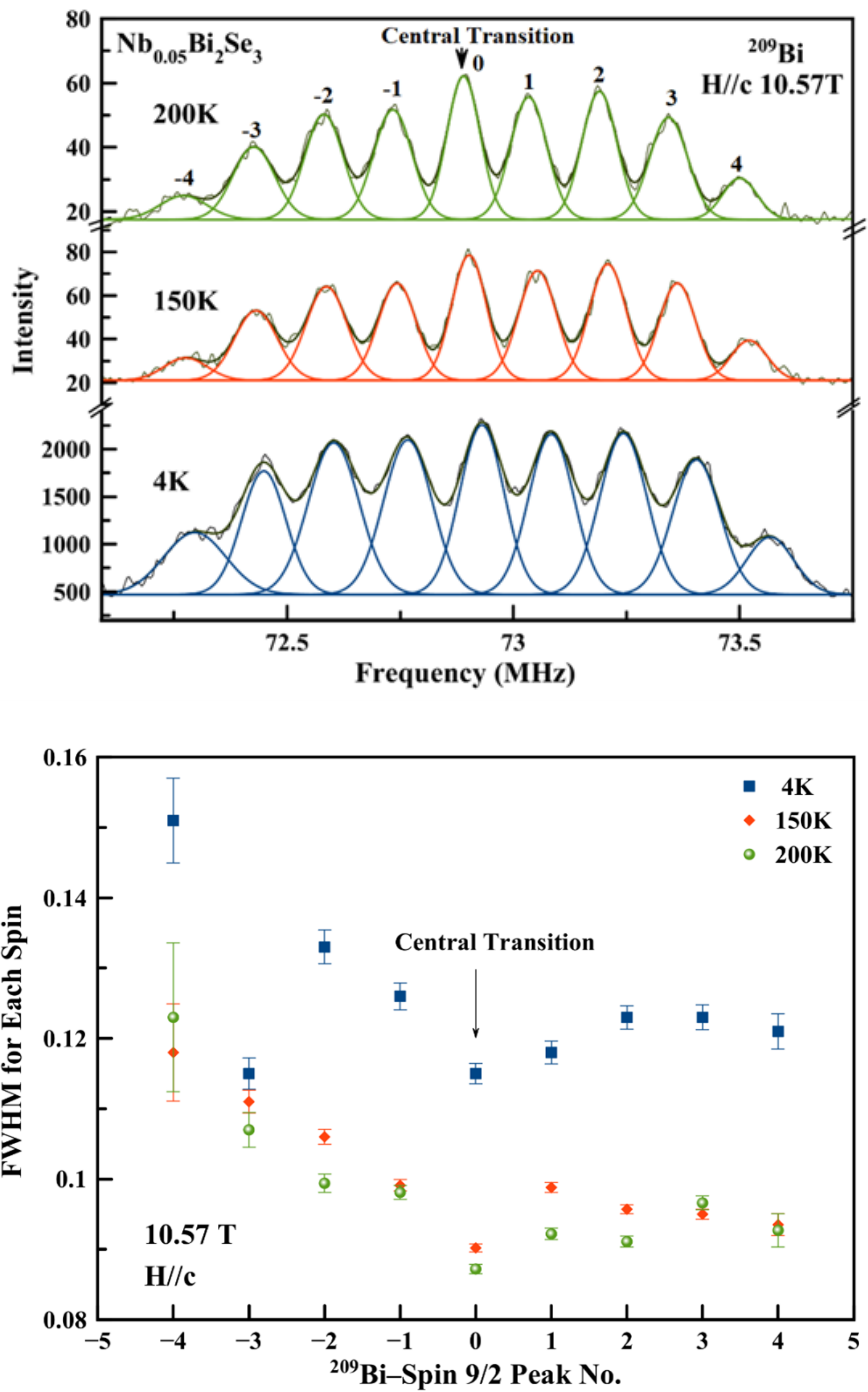

Fig 6. $^{209}$Bi NMR on crystal Nb$_{0.05}$Bi$_2$Se$_3$, with applied magnetic field H=10.57T and parallel to the c-axis of the crystal. (a) Frequency sweep spectra (72MHz to 74MHz) at temperature below T$_{CDW}$(4K, blue), near T$_{CDW}$(150K, orange) and above T$_{CDW}$(200K, green). Each peak was fitted by Gaussian function. The black curve line and dark green curve line for each spectra, represent the raw data and Fit-sum. The numbered 9 peaks are from the spin-9/2 of $^{209}$Bi, with the central peak labeled as peak 0, the satellite peaks labeled as -1, -2, -3, -4 (left) and 1, 2, 3, 4 (right) away from the center peak. (b) Full Width at Half Maximum (FWHM) for each individual peak from (a) vs Bi-Spin 9/2 Peak No. at three different temperatures. Blue square, Orange diamond and green solid ball represent 4K, 150K and 200K.





quadrupolar interaction magnitudes are in general large, where even very small local distortions can be detected.[43,44] This further leads to a change in the frequency associated with the transitions between the nuclear levels, which can be reflected as a line broadening.[45] Above the Peierls transition, a quadrupolar interaction effect can lift the degeneracy between the transitions corresponding to the different nuclear levels, given by:

$$\Delta\nu_{m\rightarrow m+1}^0(R) = \frac{2m+1}{4}\nu(Q)f(\Theta)$$

with $\nu(Q) = \frac{3e^2qQ}{2I(I-1)}$

where Q is the quadrupole moment, I the spin of the nucleus, eq the main component of the EFG tensor induced by external charges of the nuclei, and f(Θ) is orientation factor between external applied magnetic field and the EFG principal axis.

Below the Peierls transition, the electric charge is modulated and associated with the transitions between the nuclear levels is given by (Berthier and Segranson, 1987)

$$\Delta\nu_{m\rightarrow m+1}(RG) = \Delta\nu_{m\rightarrow m+1}^0(TC) + \frac{\omega_1}{2\pi}cos(2k_Fr + \varphi) + \frac{\omega_2}{2\pi}cos^2(2k_Fr + \varphi)$$

where the numerical factors $\omega_1$ and $\omega_2$ depend on the amplitude of the CDW and on the quadrupole moment. [46,47]

To discuss the satellite broadening with further away from the central transitions, it is greatly related to the underlying CDW mechanism in the compound.[41,48] Since we do not have enough evidence for the understanding of the nature of CDW (eg. How many CDW channels, dimensionality and so on) observed here, we are providing some speculations based on previous CDW related quadrupolar NMR studies. In the H//c direction, the EFG on the central transitions can be neglected.[49] The most possible reason for satellites becoming broader could be because the nature of quadrupolar splitting is an EFG induced effect and the fact that CDW can provide a supplementary EFG, thus, with experiencing a distribution of additional EFG, we observed an anomalous NMR broadening on the satellites.[48] Similar conclusion has been mentioned in Willson's dissertation and he mentioned this is an indication of discommensurate CDW, which is a mixture of ordered CDW and I-CDW coexisting.[41] Another material with Charge-Density Wave, Cuprate Superconductors,[50] has also discussed similar phenomenon where the long-range correlations below $T_{CDW}$ can produce a line shape asymmetry that grows with decreasing temperature.

In the NMR studies for line broadening near CDW phase transitions, the FWHM broadening with lowering of temperatures have been reported and discussed in the spin-9/2 [93]Nb NMR in NbSe2 and [17]O NQR(Quadrupole Satellites of NMR lines) in YBCO.[41,51] In the case of spin-9/2 [93]Nb in NbSe2,[41] CDW was signed by the FWHM broadening with lowering of temperature at central transition and the first satellites, where pre-broadening was observed and it became more rapid after $T_{CDW}$. In the case of NQR in YBCO,[51] the broadening of width starts to increase with lowering of temperature at $T_{CDW}$, in which the width is the largest with lowest available temperature. In our results, since we do not have quantities data to represent the transition shape, we presented the FWHM for individual peak at the three available temperature ranges. Similar to the above literatures' observations,[41,51] in each spin of [209]Bi as shown in Fig6 (b), FWHM at 200K is increased by 30 to 40% when the temperature drops to 4K. This homogenous broadening of FWHM between 4K to 200K further indicates the existence of CDW. Thus, our NMR from nuclear local environment further assists the possible CDW transitions that we observed from previous measurements in Nb intercalated Bi2Se3.

### 2.1.2 Material Considerations

It is important to rule out the possibility that the CDW or SC reported in this paper is from a secondary phase other than $Nb_xBi_2Se_3$, such as the well-known and existing CDW in NbSe2 or NbSe3. A detailed exploration of the published literature on these phases[52,53] reveals that the $T_c$ (superconductivity) and $T_{CDW}$ (density wave) for NbSe2 and NbSe3 are outside the range of the transitions reported in this paper. Superconducting $T_c$ for NbSe2 is 7.2K, which is outside the range of $T_c$ of Nb-intercalated Bi2Se3 reported here (1.9K to 3.4K) and elsewhere.[40] The CDW at $T_{CDW}$=33K for NbSe2 is much lower than our $T_{CDW}$ transition, which is in the range of 150-200K.[52] The other possible secondary phase, NbSe3 displays very low temperature superconductivity around 0.85 K, but under 1.5 kbar pressure. NbSe3 displays a CDW transition at $T_{CDW}$ ~145K.[53] This transition is below, but close to, the CDW transition temperatures reported here for $Nb_xBi_2Se_3$. However, we do not believe that the significant change in resistivity reported in Figure 2 can come from a minor NbSe3 secondary phase. Additionally, our observation of a CDW transition in x=0 Bi2Se3 and $T_{CDW}$ shifting with increasing Nb concentration supports our assertion that the CDW transitions we report in $Nb_xBi_2Se_3$ arise from either self-intercalation (of Bi or Se, as discussed in ref,[37] Y. Li et al and other previous papers),[54,55] or from Nb-intercalation.[20]

A recent study from Nagao et. al[56] reported SC transitions in NbBiSe3 and $Nb_{0.9}Bi_{1.2}Se_3$, where along c-axis of the crystals $T_c^{onset}$ is 3K for NbBiSe3, and it is 3.4K for $Nb_{0.9}Bi_{1.2}Se_3$. In their report, no additional CDW phase transition at higher temperature was seen in NbBiSe3, however, a hump feature between 50~150K in resistivity vs temperature was seen in $Nb_{0.9}Bi_{1.2}Se_3$. Thus, in our samples, whether CDW and SC phases are from NbBiSe3 and $Nb_{0.9}Bi_{1.2}Se_3$ needed to be further compared here. For the concerns of CDW in our samples comparing to $Nb_{0.9}Bi_{1.2}Se_3$: 1. our XRD results did not show an impurity of $Nb_{0.9}Bi_{1.2}Se_3$. 2. CDW transitions in our $Nb_xBi_2Se_3$ samples were between 160K~200K (for x=0.05 to 0.40), which are unlikely originated from an impurity phase of $Nb_{0.9}Bi_{1.2}Se_3$ (transition is at 50K~150K) but are more likely due to the different intercalation levels of Nb. For the concerns of SC in our samples comparing to $NbBiSe_3$: our samples $T_c^{onset}$ have shifts with Nb concentrations between 1.9K to 3.4K. Earlier report[40] with Nb doped $Bi_2Se_3$, where their SC phase with different Nb concentrations also have similar shifts as our samples, supports our results that SC phase is likely due to Nb intercalation and not from minor impurities. As is well known,[32,33,57] a CDW often forms together with a Periodic Lattice Distortion (PLD). A number of mechanisms can initiate a PLD: most such mechanisms arise from lattice strain. As discussed above, and elsewhere,[32,58] the fact that a lattice strain could arise in an intercalated layered compound is intuitively obvious. However, here we discuss a collection of mechanisms that could lead to a PLD in intercalated $Bi_2Se_3$.

As with other investigators,[21-22,40,59,60] our crystal growth of Nb-$Bi_2Se_3$ involved high temperature annealing and quenching. Schneeloch et al,[61] and N. P Smith,[38] studying $Cu_xBi_2Se_3$, have suggested that quenching (i.e., differences in quenching temperature, as well as the difference between quenching versus not quenching) drives the nature of intercalation, the nature of the Fermi surface, the presence or absence of superconductivity and even the superconducting fraction in a specific sample. The complete mechanism for such a critical difference on quenching in these systems remains to be fully understood. So far, our observations, combined with those of Smith ,[38] Schneeloch et al,[61] and others,[21-22,40,59,60] underscore the fact that the exact nature of intercalation and the type of electronic order in $Nb_xBi_2Se_3$ is strongly dependent upon growth and quenching. Higher temperature quenching appears to facilitate both CDW and superconductivity in these systems, but further work on $Nb_xBi_2Se_3$ is needed for us to be able to state this conclusively. Work done so far,[15,16,37,54,55,60,61] reveals that two types of intercalations can impact electronic order in $Bi_2Se_3$: (a) self-intercalation, in which Bi or Se atoms migrate from quintuple layer crystal lattice sites into intercalating spaces; and (b) intentional intercalation of external atoms such as Nb, Cu and Sr. The latter generally leads to additional availability of charge carriers,[62] and supports superconductivity. As discussed above, such intercalation can lead to 1D chains in other similar compounds[11,36,63] which supports anisotropic electron-phonon interaction along the z-axis.

It is instructive to examine the possible role of anti-phase boundaries. Using SAED and STM/STS measurements on thin films of $Bi_2Se_3$ combined with Density Functional Theory, Y Liu and others[25,26] suggested the presence of charged grain boundaries arising from anti-phase grain boundaries.[25] They explored the possibility that different nucleation sites of $Bi_2Se_3$ form, during crystal growth, along the [0001] direction. They suggest that a mismatch between the Bi-Bi plane and the Se-Se plane nucleation site grain boundaries (or the other way around) can introduce such anti-phase grain boundaries. Furthermore, if this Bi-Bi & Se-Se interface has certain repeating order (e.g. ±3/5quintuple layer variant), it will lead to charging (and potential shifts) near the grain boundaries. In our case, the $Nb_xBi_2Se_3$ are bulk crystals, and thus, the grain boundaries related mechanism could be more complicated. However, the antiphase boundaries (Bi-Bi to Se-Se or Nb-Nb) can still be formed in the as-grown crystals. With sufficient levels of such interface states, the incommensurate lattice disorder can be generated. When decreasing the temperature, the incommensurate state can be further locked into commensurate CDW state. Therefore, it is also possible that the anti-phase grain boundaries in our material can lead to CDW.

### 2.1.3 Effects of intercalation on lattice and electronic degrees of freedom

It is reasonable to argue that intercalation into a layered compound such as $Bi_2Se_3$ can lead to lattice distortion along the Γ-Z direction, further resulting in changes in the electronic and phonon structure with possible relevance to superconductivity.[17,18,64] Structural distortion along the z-direction in $Sr_{0.1}Bi_2Se_3$ has been reported by A. Yu. Kuntsevich, et al.[65] Xiangang and Sergey[17] suggest that small changes in z-direction positions of the intercalant can modify the nesting vector $X(q)=\Sigma\delta(\varepsilon_k)\delta(\varepsilon_{k+q})$, where $\varepsilon_k$ and $\varepsilon_{k+q}$ represent energy states near the Fermi level. When wavevector q is close to zero, this nesting vector along $\Gamma_Z$ is largest, possibly leading to strong Fermi nesting and strong electron-phonon coupling in the $\Gamma_Z$ direction. At small momenta (small $q_s$), displacement along [001] breaks spatial inversion symmetry and lifts double degeneracy, resulting in a large electron-phonon coupling matrix element along this direction close to the zone center. In their experiments using inelastic neutron scattering experiments on superconducting $Sr_{0.1}Bi_2Se_3$, Wang et al [18] have subsequently observed the presence of highly anisotropic phonons. The linewidths of these acoustic phonons in the [001] directions are found to increase substantially at long wavelength. The presence of an open-cylinder like electron pocket in the $\Gamma_Z$ direction,[19] centered around the Γ point, indicates the presence of Fermi surface nesting along $\Gamma_Z$, with the nesting function being strongest at low $q$. [18] It is therefore reasonable to suggest that intercalation-induced lattice distortion along [001] helps the formation of a quasi-1D Peierls-type CDW state along the z-direction of Nb-$Bi_2Se_3$, with concomitant changes in electron-phonon coupling.

Fermi nesting can result from the quasi-hexagonally deformed Fermi surface of $Bi_2Se_3$,[23,24,37] and from multiple bands and Fermi pockets. From their de Hass van Alphen study on oriented single crystals, Lawson et. al report the presence of fermi pockets in Nb-$Bi_2Se_3$ based on observation of multiple oscillation frequencies.[20] They speculate that this could be caused by the Nb d-orbital, and additionally suggest that such multiple fermi-surfaces can support the formation of a CDW state. For these and other reasons, it is not surprising that a CDW ground state exists in Nb-Bi$_2$Se$_3$. We discuss in this paper that the origin of such an effect could well be driven by the intercalation of specific atomic species (e.g., Bi, Se, Nb) rather than by the Nb d-orbital as suggested by Lawson et al. We speculate that Fermi nesting-driven density wave ground states in Bi$_2$Se$_3$ are, in general, driven by intercalation.

In Figure 4, we note a change in the trendline of $T_{CDW}$ (CDW transition temperature) and $T_c$ with increasing Nb concentration. Both $T_c$ and $T_{CDW}$ rise up to approximately x=0.20, then fall to lower values. Interestingly, this trend correlates with structural distortion, as revealed in the c/a ratio shown in supplementary data. This implies that increasing in c/a, which would cause lattice distortion, decreases upon additional intercalation. We also noticed that for our higher doped sample x=0.3 and x=0.4, the coexistence of SC and CDW is very weak and in some cases, it only appears in one case or the other, as seen in Fig.2 b, c. This might be because higher doping can suppress the CDW state as studied in other systems [66]. Li et al[32] argue that similar changes around x=0.2 in Cu$_x$Bi$_2$Te$_2$Se result from a switchover from an intercalation-only regime for x < 0.2 to a host-site substitution regime for x > 0.2. We believe that this applies to Nb$_x$Bi$_2$Se$_3$, and have discussed it further in the following section.

## CDW and SC phases in Nb-Bi$_2$Se$_3$: co-existence, competition, or cooperation?

Layered chalcogenides of different types provide a rich and varied playground in which different correlated ground states can both coexist and compete. Do the CDW and SC ground states in intercalated Bi$_2$Se$_3$ coexist cooperatively, or compete? A review of the literature on similar systems[67-72] reveals that this question is not necessarily an easy one: we anticipate that it will take a few years of effort, via multiple investigations in both theory and experiment, before drawing a well-supported mechanism. Here, having reported the presence of a CDW for the first time, we wish to begin that discussion, with a couple of examples providing possible scenarios in the context of other, previously studied layered chalcogenides.

It is generally accepted that CDW ground states tend to favor low dimensional systems.[33-35,45] This is especially the case with layered dichalcogenides such as MX$_2$ and MX$_3$, where X represents a chalcogen such as S, Se or Te, and M represents a transition metal from group IV, V or VI. [57,63] Some well-studied examples of layered chalcogenides are: MoS$_2$, TiS$_2$, TaS$_2$, WS$_2$, ZrS$_2$, WSe$_2$, Sb$_2$Se$_3$, NbSe$_2$, Bi$_2$Se$_3$ and Bi$_2$Te$_3$. These quasi-2D systems feature structures with strong in-plane bonding, but weak out-of-plane van der Waals-type interactions.[5-10,33]Increased pressure or doping/intercalation in dichalcogenides are known to induce, or enhance, superconductivity.[5,6,10,66] It is also generally accepted that the appearance of a CDW in these materials is concomitant with an electron density modulation near the Fermi surface, leading to the opening of an energy gap. (Note, however, that such a gap is not universal: in some cases, such as in 2H-NbSe$_2$, no energy gap has been observed together with a CDW ).[73, 74] Superconductivity, on the other hand, features yet another instability near the Fermi surface. Based on the idea that the CDW and SC states compete with each other in terms of availability of charge carriers and electron states near the Fermi surface, there exists a view that the two order parameters must compete in some way.[5,6,10,13,67] This is not universally observed to be the case, as discussed below.

For example, ZrTe$_3$, a quasi-2D structure, displays a CDW order at 63K which appears to compete with superconductivity. Increase of hydrostatic pressure is found to suppress $T_{CDW}$ while enhancing superconducting $T_c$.[68] With increasing doping/ intercalation with Cu or Ni, ZrTe$_3$ exhibits a coexistence of CDW and SC, [66,67] slightly suppressing CDW order while enhancing SC. In ZrTe$_{3-x}$Se$_x$, substitution of Se at the Te site results in a suppression of the CDW order and an enhancement of superconducting $T_c$.[6] In a somewhat similar story, both substitution and intercalation of Pd into IrTe$_2$ (a system with strong spin-orbital coupling), resulting in Pd$_x$IrTe$_2$ and Pd$_y$Ir$_{1-y}$Te$_2$ respectively, results in a suppression of CDW order and an enhancement of SC.[69] An in-depth review of different chalcogenides, and the effects of intercalation, substitution, doping, or pressure (chemical and/or hydrostatic), reveals that it can be difficult to generalize any relationships between CDW and SC. It is understood that intercalation tends to modify available charge carriers, which can in turn modify the nature of superconductivity. [19,75,76] The CDW state, in turn, tends to be sensitive to disorder, which tends to be modified by substitution.[56,58,69]

Layered chalcogenides also provide examples in which the CDW and SC states do not compete. In 2H-NbSe$_2$, formation of the CDW appears to "boost" SC via what is believed to be an increase in electron-phonon coupling strength.[4,70,71] In this system, disorder induced by electron irradiation yields a cooperative relationship between SC and CDW.[4] Kiss et al.[71] reveal, from temperature-dependent angle-resolved photoemission spectroscopy on 2H-NbSe$_2$ across the CDW and SC transitions ($T_{CDW}$~33 K and $T_c$=7.2 K), that a CDW-induced spectral-weight depletion at certain Fermi-surface-crossing k points evolve into the largest superconducting gaps. In such situations, the usual mean field picture might not be valid to explain CDW order. However, a sizable electron–phonon coupling is believed to be required to induce an instability, and an interplay between the CDW and SC orders of the electronic structure. A similar mechanism is believed to exist in 1H-TaSe$_2$, where the CDW and SC ground states do not appear to compete.[72]

The richness and complexity of the layered chalcogenide families makes it difficult to make any generalized statements, or to draw universal thermodynamic phase diagrams depicting the various ground states as a function of a degree of freedom (e.g., chemical pressure, doping, etc). Thus, the observable (extant) relationship between the CDW and SC ground states tends to be mired by multi-factorial parameters such as crystal structure, bonding, dimensionality, and electronic structure.

Bi$_2$Se$_3$ is a transition metal chalcogenide like the systems described above and is likely to display similar confounding multifactorial parameters. Bi$_2$Se$_3$ has a quasi-2D layered structure, with the unit cell built by Se1-Bi-Se2-Bi-Se1 quintuple layers, and adjacent layers believed to be bonded by weak Van der Waals forces. Superconductivity can be induced into Bi$_2$Se$_3$ with external pressure[77] or chemical pressure (metal intercalation). [14-16] It is believed that this unconventional superconductivity in Bi$_2$Se$_3$ is related to strong electron-phonon interaction at long wavelength.[17-19] Based on Figure 4 (a), in which we display the evolution of $T_{CDW}$ and $T_{SC}$ with varying levels of Nb-intercalation into the Van der Waals spaces, we are tempted to conclude that CDW transition $T_{CDW}$ does not compete with $T_{SC}$. As shown in Figure 4, $T_{CDW}$ increases with SC transition $T_c$ below Nb concentration x=0.20. However, for x$\geqslant$0.20, with increase in Nb concentration, the $T_{CDW}$ decreases. Rietveld refinement of X-ray diffraction [ref: supplementary material] reveals a decrease in c/a ratio, implying a change in intercalation-induced chemical pressure. Based on similar previous studies, including our own,[32] we suggest that increasing x beyond x$\geqslant$ 0.25 results in substitution of Nb into Bi sites. Additionally, we note a gradual increase in the formation of secondary phases (BiNbSe$_3$ and BiSe) with increasing x beyond 0.25, as also reported previously.[40, 59] We conclude that pushing Nb beyond x>0.25 results in increased disorder, in turn leading to a suppression of $T_{CDW}$. It is interesting to now observe that $T_{SC}$ also decreases, as if to follow $T_{CDW}$. This scenario of disorder-induced suppression of $T_{CDW}$ also appears to be borne out from the broadening of the $T_{CDW}$ transition for x$\geqslant$0.25 shown in Figure 3a, eventually smeared out by the time we reach a Nb concentration of x=0.4. Thus, a competition between CDW and SC were not seen in our Nb$_x$Bi$_2$Se$_3$ samples. For whether there is a coexistence of SC and CDW states below $T_C$, low temperature diffraction, neutron studies, and ARPES could clarify this question further. We will venture to claim that, in Nb-Bi$_2$Se$_3$, the CDW and SC are likely generated from chemical doping effect.

## 3. Summary

In summary, we have revealed the existence of Charge density Wave order in Nb intercalated Bi$_2$Se$_3$. We further report that CDW in this system is accompanied with superconductivity in all our samples of Nb$_x$Bi$_2$Se$_3$, 0 < x < 0.4. In the 0$\leq$ x$\leq$0.20 regime, $T_{CDW}$ and $T_C$ increase with increasing Nb concentration. In the x>0.20 regime, both $T_{CDW}$ and $T_C$ decrease with increasing Nb concentration. This is possibly due to a disorder-induced suppression of both charge orders. Further work is required in order to understand whether the two orders coexist or compete at low temperature. Using Selected Area Electron Diffraction (SAED), which shows diffuse "streaks" and satellite reflections, we confirm the presence of a Periodic Lattice Distortion (PLD) and an Incommensurate Charge Density Wave (I-CDW) at room temperature. Our temperature dependent resistivity measurements reveal metal-to-insulator transitions from 140K to 200K. On sample Nb$_{0.05}$Bi$_2$Se$_3$, high field [209]Bi NMR spectra line shape with broadening at 4K (below $T_{CDW}$) was compared with spectra line shape at 200K/255K (above $T_{CDW}$), which further confirmed our locked-in CDW state. We believe that a higher-temperature I-CDW state locks into a CDW ordered state around 150-200K, which we assign as the CDW transition temperature $T_{CDW}$. We discuss our results in the context of weak coupling between layers, and distortion-induced charge order along the z-axis. We also discuss the role of strong Fermi nesting, electron-phonon interaction, spin-orbital coupling, and multiple band structures. This paper reveals that a better understanding of the charge order revealed here is important to the eventual understanding of superconductivity induced in this quantum material which is also a topological insulator. Further measurements, and theoretical examination will be needed in order to fully reveal the underlying physics in Nb-Bi$_2$Se$_3$.

## Experimental Methods

Nb$_x$Bi$_2$Se$_3$ single crystals with 0$\leq$x $\leq$ 0.4 were grown by self-flux method. High-purity (99.999%) powders of Bi, Se and (99.99%) Nb in stoichiometric ratios were prepared by using a method similar to that described in ref.[56] Stoichiometric mixtures of 2.5g batches were sealed into high-quality quartz tubes in vacuum after being weighed and sealed in an inert glove box, taking care to never expose to air. The mixtures in sealed quartz tubes were heated up to 850°C and maintained at that temperature for 20 hours. They were then cooled to 650°C at 0.1°C/min, followed by quenching into ice water. Crystals with x > 0.4 are of relatively poor quality and are not discussed here.

Powder X-ray Diffraction (XRD) measurements were performed using a Bruker D8 Discover x-ray diffractometer with Cu Kα radiation. Rietveld refinement was performed using GSAS (General Structure Analysis System) with an EXPGUI interface. Selected Area Diffraction (SAED) was performed at room temperature with a Hitachi H-9000NAR high-resolution transmission electron microscope (HRTEM) operated at 300 kV. Crystals of Nb$_x$Bi$_2$Se$_3$ (x= 0.00, 0.05, 0.15, 0.20 and 0.25) were gently ground by hand in a mortar and pestle and dispersed on lacey-carbon grid. Variable temperature resistivity studies were performed using 4-probe silver paste contacts on single crystals placed in a Quantum Design (QD) 9 Tesla Physical Property Measurement System (PPMS), where measurements were performed along J//c-axis of the crystals. Temperature and field-dependent magnetization measurements were performed using a QD MPMSXL-5 Magnetic Property Measurement System (MPMS) using a superconducting quantum interference device (SQUID). Pulsed [209]Bi Nuclear Magnetic Resonance (NMR) was performed on $Nb_{0.05}Bi_2Se_3$ single crystal of crystal size ~5.3 x 4.2 x 1.2 mm placed inside a home-built probe in an 18-Tesla Helium cryostat. The single crystal $Nb_{0.05}Bi_2Se_3$ was studied with magnetic field oriented in H//c-axis, where the orientation was determined by angle-dependent measurements. Spin-echo signals for $^{209}$Bi NMR spectra were processed using the summed Fourier transform method, with frequency swept from 72MHz to 74MHz and static magnetic field H=10.57T.

**Author contributions**:

YL and CP grew samples. Electron diffraction was performed by YL and MAS, led by MAS. YL performed the resistivity and DC magnetic susceptibility measurements. YL, CP, SR and AD performed NMR measurements under the direction of AR; CP and AD assisted YL in analysis. YL wrote the first draft of the manuscript under the direction of PG, with assistance from AR, CP and AD and reviewed by AR and RB. PG provided supervision of the overall project.

**Acknowledgment:**

A portion of this work was performed at the National High Magnetic Field Laboratory, which is supported by the National Science Foundation Cooperative Agreement No. DMR-1644779 and the State of Florida. RB acknowledges support from the National Science Foundation through NSF/DMR1904361.

**Reference**

1. Isobe, H., Yuan, N. F. Q. & Fu, L. Unconventional Superconductivity and Density Waves in Twisted Bilayer Graphene. *Physical Review X* **8**, (2018).

2. Cho, D. N., Brink, J. van den, Fehske, H., Becker, K. W. & Sykora, S. Unconventional superconductivity and interaction induced Fermi surface reconstruction in the two-dimensional Edwards model. *Scientific Reports* **6**, (2016).

3. Lee, S. *et al.* Unconventional charge density wave order in the pnictide superconductor Ba(Ni1-xCox)2As2. *Physical Review Letters* **122**, (2019).

4. Cho, K. *et al.* Using controlled disorder to probe the interplay between charge order and superconductivity in NbSe2. *Nature Communications* **9**, (2018).

5. Li, L. *et al.* Superconducting order from disorder in 2H-TaSe 2- x S x. *npj Quantum Materials* **2**, (2017).

6. Zhu, X. *et al.* Superconductivity and Charge Density Wave in ZrTe3-xSex. *Scientific Reports* **6**, (2016).

7. Wang, B. *et al.* Universal phase diagram of superconductivity and charge density wave versus high hydrostatic pressure in pure and Se-doped 1T-Ta S2. *Physical Review B* **97**, (2018).

8. Harper, J. M. E., Geballe, T. H. & Disalvo, F. J. Therma properties of layered transition-metal dichalcogenides at charge-density-wave transitions*. *Physical Review B* **15**, (1977).

9. Moncton D. E., Axe J. D. & DiSalvo F. J. Study of Superlattice Formation in 2H-NbSe2 and 2H-TaSe2 by Neutron Scattering. *Physical Review Letters* **34**, (1975).

10. Fang, L. *et al.* Fabrication and superconductivity of NaxTaS2 crystals. *Physical Review B - Condensed Matter and Materials Physics* **72**, (2005).

11. Denholme, S. J. *et al.* Coexistence of superconductivity and charge-density wave in the quasi-one-dimensional material HfTe3. *Scientific Reports* **7**, (2017).

12. Weber, F. *et al.* Extended phonon collapse and the origin of the charge-density wave in 2H-NbSe2. *Physical Review Letters* **107**, (2011).

13. Suderow, H., Tissen, V. G., Brison, J. P., Martínez, J. L. & Vieira, S. Pressure induced effects on the fermi surface of superconducting 2H-NbSe2. *Physical Review Letters* **95**, (2005).

14. Sasaki, S. *et al.* Topological superconductivity in CuxBi2Se3. *Physical Review Letters* **107**, (2011).

15. Asaba, T. *et al.* Rotational symmetry breaking in a trigonal superconductor Nb-doped Bi2Se3. *Physical Review X* **7**, (2017).

16. Shruti, Maurya, V. K., Neha, P., Srivastava, P. & Patnaik, S. Superconductivity by Sr intercalation in the layered topological insulator Bi2Se3. *Physical Review B - Condensed Matter and Materials Physics* **92**, (2015).

17. Wan, X. & Savrasov, S. Y. Turning a band insulator into an exotic superconductor. *Nature Communications* **5**, (2014).

18. Wang, J. *et al.* Evidence for singular-phonon-induced nematic superconductivity in a topological superconductor candidate Sr0.1Bi2Se3. *Nature Communications* **10**, (2019).

19. Lahoud, E. *et al.* Evolution of the Fermi surface of a doped topological insulator with carrier concentration. *Physical Review B - Condensed Matter and Materials Physics* **88**, (2013).

20. Lawson, B. J. *et al.* Multiple Fermi surfaces in superconducting Nb-doped Bi2Se3. *Physical Review B* **94**, (2016).

21. Shen, J. *et al.* Nematic topological superconducting phase in Nb-doped Bi2Se3. *npj Quantum Materials* **2**, (2017).

22. Cho, C.-W. *et al. Z3-vestigial nematic order due to superconducting fluctuations in the doped topological insulators NbxBi2Se3 and CuxBi2Se3.*

23. Fu, L. Hexagonal warping effects in the surface states of the topological insulator Bi2Te3. *Physical Review Letters* **103**, (2009).

24. Kuroda, K. *et al.* Hexagonally deformed fermi surface of the 3D topological insulator Bi 2Se3. *Physical Review Letters* **105**, (2010).

25. Liu, Y. *et al.* Charging dirac states at antiphase domain boundaries in the three-dimensional topological insulator Bi2Se3. *Physical Review Letters* **110**, (2013).

26. Liu, Y. *et al.* Tuning dirac states by strain in the topological insulator bi 2 se 3. *Nature Physics* **10**, 294–299 (2014).

27. Koski, K. J. *et al.* Chemical intercalation of zerovalent metals into 2D layered Bi 2Se 3 nanoribbons. *Journal of the American Chemical Society* **134**, 13773–13779 (2012).

28. Smith, N. *Crystal Growth and Manipulation of Intercalated Chalcogenides as Superconductors and Topological Insulators*. https://dc.uwm.edu/etd/1921 (2018).

29. Matano, K. et al. Spin-rotation symmetry breaking in the superconducting state of Cu$x$Bi2Se3 . *Nat. Phys.* **12**, 852–854 (2016).

30. Yonezawa, S. Nematic superconductivity in doped Bi$_2$Se$_3$ topological superconductors. *Condens. Matter* **4**, 2 (2019).

31. Hecker, M. & Schmalian, J. Vestigial nematic order and superconductivity in the doped topological insulator Cu x Bi2Se3. *npj Quantum Materials* **3**, (2018).

32. Li, Y., Smith, N. P., Rexhausen, W., Schofield, M. A. & Guptasarma, P. Possible lattice and charge order in Cu x Bi2Te2Se . *Journal of Physics: Materials* **3**, 015008 (2019).

33. Wilson, J. A., Salvo, F. J. di & Mahajan, S. Charge-Density Waves in Metallic, Layered, Transition-Metal Dichalcogenides. *Physical Review Letters* **32**, 16 (1974).

34. Williams, P. *et al.* CHARGE DENSITY WAVES IN THE LAYERED TRANSITION METAL DICHALCOGENIDES. *Journal de Physique Colloques* **37**, (1976).

35. Mcmillan, W. L. *Landau theory of charge-density waves in transition-metal dichalcogenides\*. PHYSICAL REVI EW B* vol. 12.

36. Zhang, H. et al. Topological insulators in Bi2Se3, Bi2Te3 and Sb2Te3 with a single Dirac cone on the surface. *Nature Phys.* **5**, 438–442 (2009).

37. Li, Y. *et al. Charge Density Wave Order in the Topological Insulator Bi 2 Se 3.*

38. Wang, M. & Koski, K. J. Polytypic phase transitions in metal intercalated Bi2Se3. *Journal of Physics Condensed Matter* **28**, (2016).

39. di Salvo, F. J. & Rice, T. M. Charge-Density Waves in Transition-Metal Compounds. *Physics Today* **32**, 32–38 (1979).

40. Kobayashi, K., Ueno, T., Fujiwara, H., Yokoya, T. & Akimitsu, J. Unusual upper critical field behavior in Nb-doped bismuth selenides. *Physical Review B* **95**, (2017).

41. Douglas M. (Douglas Mark) Wilson, 2017, NMR Investigation of the Layered Superconductor NbSe2, Florida State University

42. Lloyd L. (Lloyd Laporca) Lumata, 2008, Spin Dynamics of Density Wave and Frustrated Spin Systems Probed by Nuclear Magnetic Resonance, Florida State University

43. NMR Studies of Phase Transitions C. ODIN Groupe Matie`re Condense´e et Mate´riaux, UMR6626 au CNRS, Universite´ Rennes I,Campus de Beaulieu Bat11A, 35042 Rennes Cedex, France

44. R. J. Darton, P. Wormald and R. E. Morris, J. Mater. Chem., 2004, 14, 2036–2040.

45. Gruner, G. The dynamics of charge-density waves. *Rev. Mod. Phys.* **60**, 1129-undefined (1988).

46. *Berthier* and P. *Segranson*, "NMR Studies of *Charge Density waves* in low Dimensional Conductors", P. 455. *1987*, Vol.168. P455

47. L.P. Gor'kov, G. Grüner, "Charge Density Waves in Solids", 2000, P.188

48. Robert M. White, Theodore H. Geballe, 1984, Long Range Order in Solids: Solid State Physics, P. 197

49. J.C. Phillips & M.F. Thorpe, 2001, Phase Transitions and Self-Organization in Electronic and Molecular Networks, P.415

50. Wu, T. et al. Incipient charge order observed by NMR in the normal state of YBa$_2$Cu$_3$O$y$ *Nat. Commun.* **6**, 6438 (2015).

51. Kharkov, Y., Sushkov, O. The amplitudes and the structure of the charge density wave in YBCO. *Sci Rep* **6**, 34551 (2016).

52. Lian, C. S., Si, C. & Duan, W. Unveiling Charge-Density Wave, Superconductivity, and Their Competitive Nature in Two-Dimensional NbSe2. *Nano Letters* **18**, 2924–2929 (2018).

53. Monceau, P., Peyrard, J., Richard, J & Molinie, P. Superconductivity of the Linear Trichalcogenide NbSe3 under Pressure. *Physical Review Letters* **39**, (1977).

54. Huang, F. T. *et al.* Nonstoichiometric doping and Bi antisite defect in single crystal Bi 2Se 3. *Physical Review B - Condensed Matter and Materials Physics* **86**, (2012).

55. Cermak, P. *et al.* High power factor and mobility of single crystals of Bi2Se3 induced by Mo doping. *Journal of Solid State Chemistry* **277**, 819–825 (2019).

56. Solid State Communications 321 (2020) 114051

57. Rossnagel, K. On the origin of charge-density waves in select layered transition-metal dichalcogenides. *Journal of Physics Condensed Matter* vol. 23 (2011).

58. Zhang, F. *et al.* Lattice distortion and phase stability of Pd-Doped NiCoFeCr solid-solution alloys. *Entropy* **20**, (2018).

59. Qiu, Y. et al. Time reversal symmetry breaking superconductivity in topological materials.

60. Wang, J. *et al.* Investigate the Nb doping position and its relationship with bulk topological superconductivity in NbxBi2Se3 by X-ray photoelectron spectra. *Journal of Physics and Chemistry of Solids* **137**, (2020).

61. Schneeloch, J. A., Zhong, R. D., Xu, Z. J., Gu, G. D. & Tranquada, J. M. Dependence of superconductivity in CuxBi2Se3 on quenching conditions. *Phys. Rev. B* **91**, (2015).

62. Mazumder, K., Sharma, A., Kumar, Y. & Shirage, P. M. Effect of Cu intercalation on humidity sensing properties of Bi2Se3 topological insulator single crystals. *Physical Chemistry Chemical Physics* **20**, 28257–28266 (2018).

63. Chen, C. W., Choe, J. & Morosan, E. Charge density waves in strongly correlated electron systems. *Reports on Progress in Physics* **79**, (2016).

64. Brydon, P. M. R., das Sarma, S., Hui, H. Y. & Sau, J. D. Odd-parity superconductivity from phonon-mediated pairing: Application to Cux Bi2 Se3. *Physical Review B - Condensed Matter and Materials Physics* **90**, (2014).

65. Kuntsevich, A. Y. *et al.* Structural distortion behind the nematic superconductivity in SrxBi2Se3. *New Journal of Physics* **20**, (2018).

66. Zhu, X., Lei, H. & Petrovic, C. Coexistence of bulk superconductivity and charge density wave in Cu xZrTe3. *Physical Review Letters* **106**, (2011).

67. Lei, H., Zhu, X. & Petrovic, C. Raising Tc in charge density wave superconductor ZrTe 3 by Ni intercalation. *EPL* **95**, (2011).

68. Yomo, R., Yamaya, K., Abliz, M., Hedo, M. & Uwatoko, Y. Pressure effect on competition between charge density wave and superconductivity in ZrTe3: Appearance of pressure-induced reentrant superconductivity. *Physical Review B - Condensed Matter and Materials Physics* **71**, (2005).

69. Yang, J. J. *et al.* Charge-orbital density wave and superconductivity in the strong spin-orbit coupled IrTe 2 Pd. *Physical Review Letters* **108**, (2012).

70. Xi, X. *et al.* Strongly enhanced charge-density-wave order in monolayer NbSe 2. *Nature Nanotechnology* **10**, 765–769 (2015).

71. Kiss, T. *et al.* Charge-order-maximized momentum-dependent superconductivity. *Nature Physics* **3**, 720–725 (2007).

72. Lian, C. S. *et al.* Coexistence of Superconductivity with Enhanced Charge Density Wave Order in the Two-Dimensional Limit of TaSe2. *Journal of Physical Chemistry Letters* **10**, 4076–4081 (2019).

73. Straub, T. *et al.* Charge-Density-Wave Mechanism in 2H-NbSe 2 : Photoemission Results. *Physical Review Letters* **82**, (1999).

74. Harper, J. M. E., Geballe, T. H. & Salvo, F. J. di. HEAT CAPACITY OF 2H-NbSe 2 AT THE CHARGE DENSITY WAVE TRANSITION *. *Physics Letters* **54A**, (1975).

75. Kawasugi, Y. *et al.* Electron-hole doping asymmetry of Fermi surface reconstructed in a simple Mott insulator. *Nature Communications* **7**, (2016).

76. Ishii, K. *et al.* Momentum dependence of charge excitations in the electron-doped superconductor Nd1.85Ce0.15CuO4: A resonant inelastic x-ray scattering study. *Physical Review Letters* **94**, (2005).

77. Kirshenbaum, K. *et al.* Pressure-induced unconventional superconducting phase in the topological insulator Bi2Se3. *Physical Review Letters* **111**, (2013).